\documentclass[iop,revtex4]{emulateapj}

\usepackage{natbib}
\usepackage{lscape}
\usepackage{rotating}
\usepackage{adjustbox}
\usepackage{amsmath}
\usepackage{hyperref}
\usepackage{cleveref}
\usepackage{marginnote}
\usepackage{threeparttable}
\usepackage{comment}
\hypersetup{colorlinks=true, linkcolor=blue, citecolor=blue}

\newcommand{\SFRIR}{SFR$_{\text{IR}}$ }

\crefformat{figure}{#2{\color{blue}#1}#3}
\Crefformat{figure}{#2{\color{blue}#1}#3}

\crefformat{equation}{#2{\bfseries\color{blue}#1}#3}
\Crefformat{equation}{#2{\itshape\color{blue}#1}#3}

\shorttitle{\footnotesize Near-IR emission lines in starburst galaxies at \lowercase{$0.5<z<0.9$}: Discovery of a merger sequence of extreme obscurations}
\shortauthors{A.Calabr\`o et al.}

\begin{document}


\title{Near-infrared emission lines in starburst galaxies at \lowercase{$0.5<z<0.9$}:\\ Discovery of a merger sequence of extreme obscurations}


\author{A. Calabr{\`o}\altaffilmark{1}} 
\author{E. Daddi\altaffilmark{1}}
\author{P. Cassata\altaffilmark{2}}
\author{M. Onodera\altaffilmark{3,4}}
\author{R. Gobat\altaffilmark{5}}
\author{A. Puglisi\altaffilmark{1}}
\author{S. Jin\altaffilmark{1,6}}
\author{D. Liu\altaffilmark{7}}
\author{R. Amor{\'i}n\altaffilmark{8,9}}
\author{N. Arimoto\altaffilmark{10}}
\author{M. Boquien\altaffilmark{11}}
\author{R. Carraro\altaffilmark{12}}
\author{D. Elbaz\altaffilmark{1}}
\author{E. Ibar\altaffilmark{12}}
\author{S. Juneau\altaffilmark{13}}
\author{F. Mannucci\altaffilmark{14}}
\author{H. M{\'e}ndez Hern{\'a}nez\altaffilmark{12}}
\author{E. Oliva\altaffilmark{14}}
\author{G. Rodighiero\altaffilmark{2}}
\author{F. Valentino\altaffilmark{15,16}}
\author{A. Zanella\altaffilmark{17}}

\affiliation{1, CEA, IRFU, DAp, AIM, Universit{\'e} Paris-Saclay, Universit{\'e} Paris Diderot, Sorbonne Paris Cit{\'e}, CNRS, F-91191 Gif-sur-Yvette, France
\\2, Dipartimento di Fisica e Astronomia “G. Galilei”, Universit{\`a} di Padova, Vicolo dell'Osservatorio 3, 35122, Italy
\\3, Subaru Telescope, National Astronomical Observatory of Japan, National Institutes of Natural Sciences (NINS), 650 North A'ohoku Place, Hilo, HI, 96720, USA
\\4, Department of Astronomical Science, SOKENDAI (The Graduate University for Advanced Studies), 650 North A'ohoku Place, Hilo, HI, 96720, USA 
\\5, Instituto de F{\'i}sica, Pontificia Universidad Cat{\'o}lica de Valpara{\'i}so, Casilla 4059, Valpara{\'i}so, Chile  
\\6, Key Laboratory of Modern Astronomy and Astrophysics in Ministry of Education, School of Astronomy and Space Science, Nanjing University, Nanjing 210093, China
\\7, Max Planck Institute for Astronomy, Konigstuhl 17, D-69117 Heidelberg, Germany
\\8, Cavendish Laboratory, University of Cambridge, 19 JJ Thomson Avenue, Cambridge, CB3 0HE, UK.
\\9, Kavli Institute for Cosmology, University of Cambridge, Madingley Road, Cambridge CB3 0HA, UK
\\10, Astronomy Program, Department of Physics and Astronomy, Seoul National University, 599 Gwanak-ro, Gwanak-gu, Seoul, 151-742, Korea
\\11, Centro de Astronom{\'i}a (CITEVA), Universidad de Antofagasta, Avenida Angamos 601, Antofagasta, Chile 
\\12, Instituto de Fisica y Astronom{\'i}a, Facultad de Ciencias, Universidad de Valpara{\'i}so, Gran Breta{\~n}a 1111, Playa Ancha, Valpara{\'i}so, Chile
\\13, National Optical Astronomy Observatory, 950 N. Cherry Avenue, Tucson, AZ 85719, USA
\\14, INAF-Osservatorio Astrofisico di Arcetri, Largo Enrico Fermi 5, 50125 Firenze, Italy
\\15, Dawn Cosmic Center, Niels Bohr Institute, University of Copenhagen Juliane Maries Vej 30, DK-2100 Copenhagen, Denmark
\\16, Dark Cosmology Centre, Niels Bohr Institute, University of Copenhagen, Juliane Maries Vej 30, DK-2100 Copenhagen, Denmark
\\17, European Southern Observatory, Karl Schwarzschild Stra\ss e 2, 85748 Garching, Germany
}

\begin{abstract}
We obtained optical/near-IR rest-frame Magellan FIRE spectra (including Pa$\beta$ and Pa$\gamma$) 
of 25 starburst galaxies at $0.5<z<0.9$, with average star formation rates (SFR) $\times7$ above the  Main Sequence (MS).
We find that Paschen-to-Balmer line ratios saturate around a constant value corresponding to $A_{\rm V}\sim2$--3~mag,
 while line to IR luminosity ratios suggest a large range of more extreme obscurations
 and appear to be uncorrelated to the former. 
This behavior is not consistent with standard attenuation laws derived for local and distant galaxies, while being remarkably
consistent with observations of starburst cores in which young stars and dust are homogeneously mixed. This model implies $A_{\rm V}=2$--30~mag attenuation to the center of starburst cores, with a median of $\sim9$~mag (a factor of $4000$). 
X-ray hardness ratios for 6 AGNs in our sample and  column densities derived from observed dust masses and radio sizes independently confirm this level of attenuation. 
In these conditions observed optical/near-IR emission  comes from surface regions, while inner starburst cores are invisible. We thus 
attribute the high [NII]/H$\alpha$ ratios to widespread shocks from accretion, turbulence and dynamic disturbances rather than to AGNs. 
The large range of optical depths demonstrates that substantial diversity is present within the starburst population, possibly connected to different merger phases or progenitor properties. The majority of our targets are, in fact, morphologically classified as mergers.  We argue that the extreme obscuration provides in itself  smoking gun evidence of their merger origin, and  a powerful tool for identifying mergers at even higher redshifts.
\end{abstract}


\section{Introduction}

Starburst galaxies (SBs) outliers from the MS \citep[e.g.,][]{noeske07,daddi07}, might be key to understand a long-standing mystery in galaxy formation and evolution: the transition from star forming galaxies to massive, passively evolving ellipticals. According to a popular scenario \citep[e.g.,][]{dimatteo05,hopkins10}, this transition is attributed to major mergers producing strong bursts of star formation in very dense cores and triggering obscured black hole accretion, which can both remove the gas and dust content in the galaxy.

Local ultra-luminous infrared galaxies (ULIRGs) are showcase examples of merger-induced starbursts, showing compact and heavily obscured cores \citep[e.g.,][]{soifer00,juneau09}, in agreement with the above scenario. Using standard attenuation recipes in ULIRGs \citep[][]{cardelli89,calzetti00} leads to UV, optical and near-IR based SFRs being systematically underestimated compared to the total infrared luminosities, implying optically thick conditions for these tracers \citep[][]{goldader02,garcia09,rieke09}.

\begin{figure*}[ht]
  \centering
  \includegraphics[angle=0,width=18.5cm,trim={3cm 0.cm 3cm 0cm},clip]{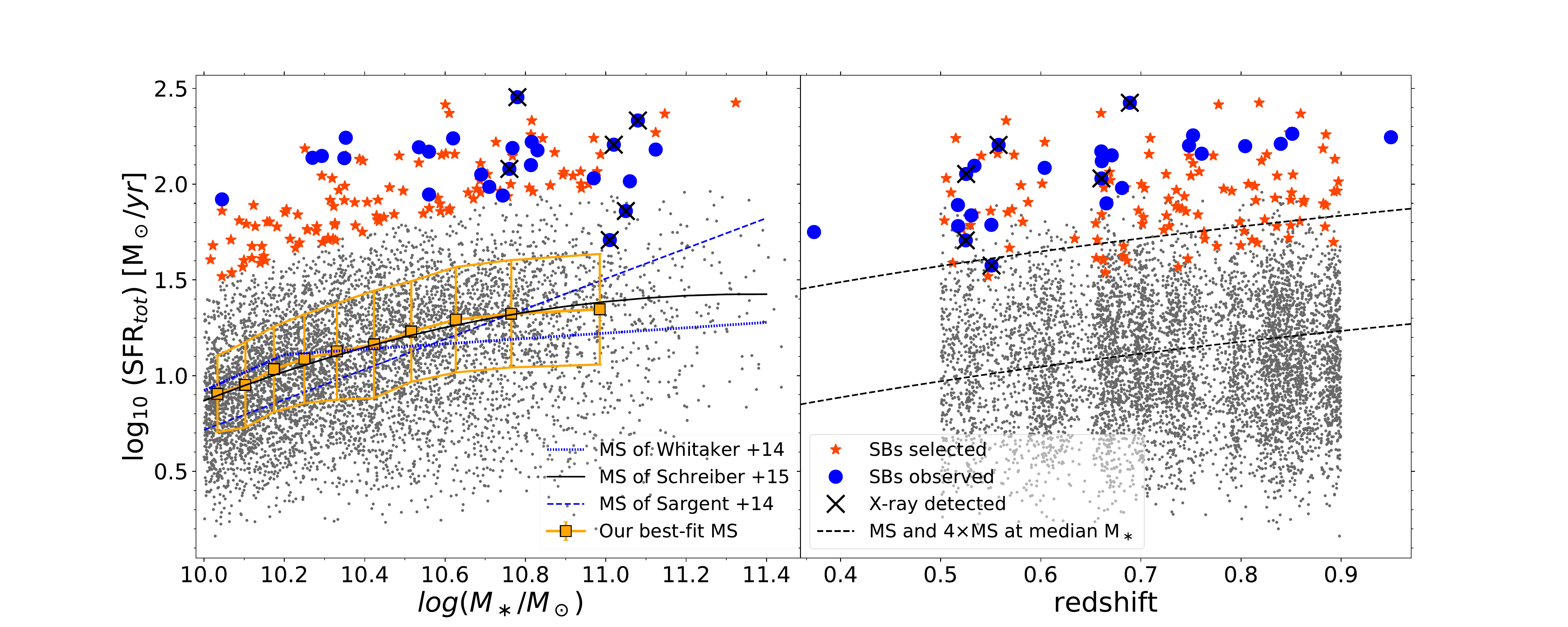}
  \caption{\small \textit{Left:} SFR-M$\star$ diagram for galaxies in our parent sample ($0.5<$z$<0.9$) , where SFR$_\text{tot}$ is defined as SFR$_\text{UV,obs}$+SFR$_\text{IR}$, and is normalized to their median redshift ($0.73$) using the evolving trend from \citet{sargent14}. For sources detected only at 24$\mu$m we estimated the SFR from their $24\mu$m flux using M12 templates. 
  \textit{Right:} SFR$_{\text{tot}}$ vs redshift for the same sample. 
  }\label{selection}
\end{figure*}


The nature and evolution of SBs galaxies in the distant Universe is debated. While they might still be major merger events, there are also claims that they might be instead very gas rich galaxies \citep[e.g.,][]{scoville16}, possibly due to exceptionally strong gas accretion events. This is supported by ideas that at higher redshifts, with higher gas fractions, major mergers might only rarely result in strong SFR enhancements \citep{fensch17}. 
Comparisons of both dust-free and dust-affected SFRs are required to study their degree of obscuration, providing clues on the ULIRGs/distant-SB connection.
\citet{puglisi17} showed that on average Balmer emission lines of Herschel-selected  $z\sim1.6$ SBs are mainly coming from  regions producing $<10 \%$ of the total SFR, suggesting that rest-optical lines cannot be used to infer the physical properties of the whole starburst system. 
These results prompted us to use Magellan FIRE to obtain spectroscopy of starbursts in the near-IR rest-frame 
with the aim of providing enhanced sensitivity and constraining power to study their attenuation properties (hence their nature).  In this letter we present first results of this effort. We adopt \citet{chabrier03} IMF, AB magnitudes and standard cosmology ($H_{0}=70$ $\rm km s^{-1}Mpc^{-1}$, $\Omega_{\rm m} = 0.3$, $\Omega_\Lambda = 0.7$).

\section{Sample selection}\label{second}

We select starbursts galaxies for observations with Magellan FIRE in the COSMOS field, with the following criteria:

\begin{itemize}\raggedright
    \item spectroscopic redshift  $0.5<$ z $<0.9$\footnote{We also added four galaxies with photometric redshifts \citep{laigle16} lying in the same range.} (from optical surveys, Salvato et al. in preparation), placing Pa$\beta$  within the K band, and H$\alpha$  above $0.82 \mu m$, thus observable with FIRE\footnote{Existing spectroscopic redshift were incorrect for two galaxies, new redshifts placing them outside of our selection range (Fig. \cref{selection}-\textit{right}). We keep them in the sample as they satisfy the other selection criteria.}.
    \item SFR $> 4 \times$ SFR$_{\text{MS}}$ \citep{rodighiero11}\footnote{Because of the unavoidable variations of the IR photometry  among different catalog versions, while they were being built, two objects appear now slightly below this $\times4$ threshold (Fig. \cref{selection}-\textit{left}).}. SFRs are derived using the IR catalog from \citet{jin18}\footnote{Accurate SFR measurements were derived by fitting IRAC to radio $20$cm photometry from \citet{jin18} with four components as follows: a \citet{bruzual03} SED for the stellar component (with age $200$ Myr, constant SFH, Z$_\odot$, Chabrier IMF and Calzetti attenuation law), a mid-infrared AGN template from \citet{mullaney11} and a warm+cold dust SED from the full \citet{draine07} library.}. As shown in Fig. \cref{selection}, the MS for our sample, derived through a running median over $10$ bins in M$_\star$, agrees with the literature \citep[][]{sargent14,schreiber15}. 
    Our SFRs are de-contaminated from AGN torus emission ($3\%$ median contribution to L$_{\rm IR}$; see \citet{liu18} and \citet{jin18} for the procedure). 
    \item M$_\ast$ $> 10^{10}$ M$_\odot$, for sample completeness: above this mass limit and up to  $z=0.9$, all SBs would be Herschel detected at S/N$_{\text{FIR}}>5$ (cf. Fig.~16 in \citet{jin18}). Stellar masses are from \citet{laigle16}.
 
\end{itemize}

These criteria yield a total of 152 starburst candidates for our Magellan observations (Fig. \cref{selection}). They represent $2$-$3\%$ of the whole star-forming population in the same mass range and redshift, 
\citep[see, e.g.,][]{sargent12,sargent14,schreiber15}.

\section{Magellan-FIRE observations}\label{third}

\begin{figure*}[t!]
  \centering
  \includegraphics[angle=0,width=18cm,trim={0cm 0cm 0cm 0.cm},clip]{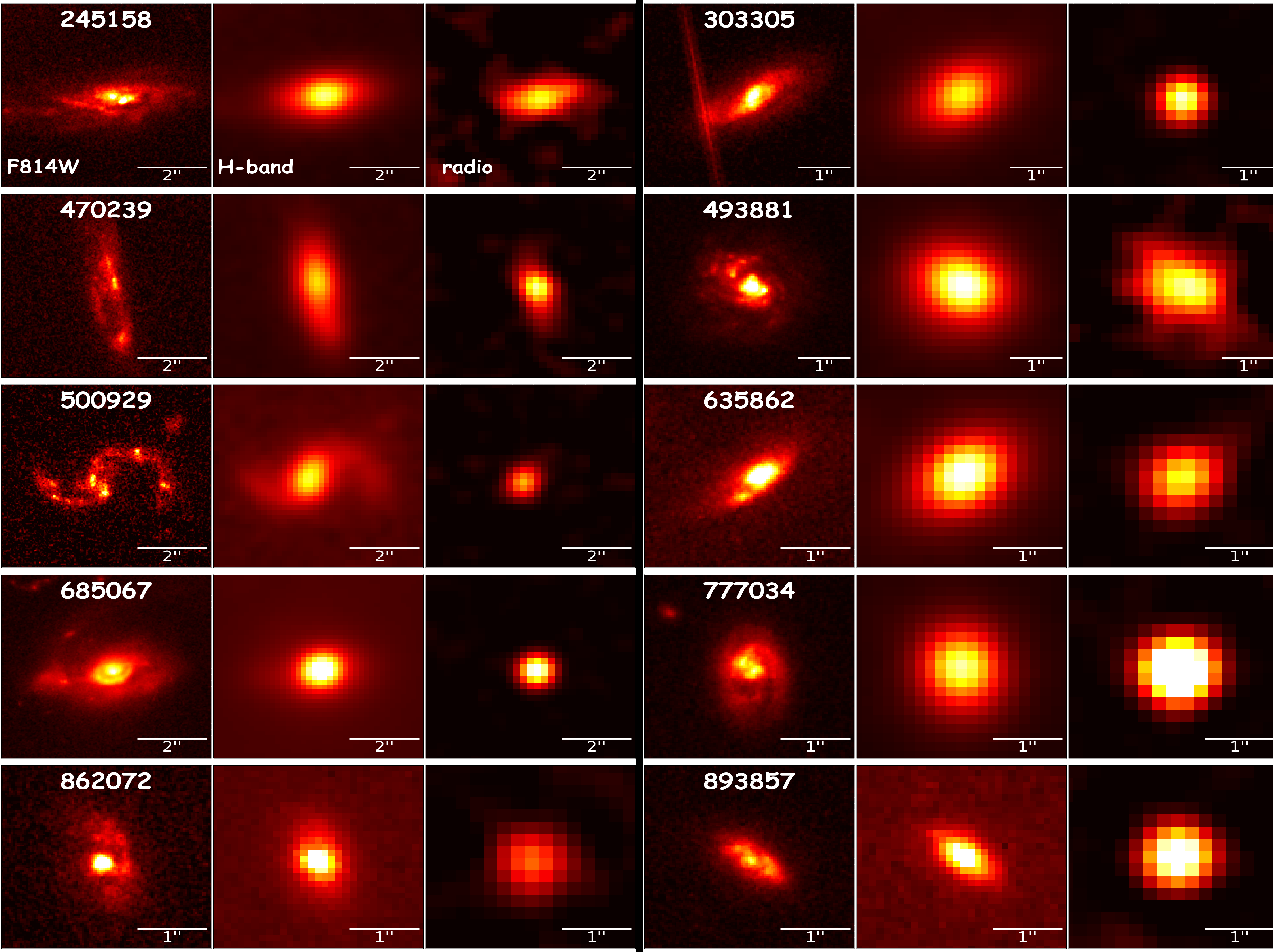}
  \caption{\small \textit{Left:} HST-ACS F814W cutout images for $10$ galaxies of our Magellan sample. Their FWHM resolution is $0.095''$.  \textit{Center:} H-band UltraVISTA images with the same f.o.v. and with FWHM$_{res} \sim 0.75''$. \textit{Right:} 3 GHz radio images from VLA-COSMOS 3GHz Large Project \citep{smolcic17}, FWHM$\sim 0.75''$. 
  For the galaxies $862072$ and $893857$, higher resolution ($0.2''$) H-band cutouts from the COSMOS-DASH program are shown  \citep{momcheva16}.}\label{cutouts}
\end{figure*}

\begin{figure*}[t!]
  \hspace*{-2em} \centering
  \includegraphics[angle=0,height=4.9cm,trim={0.5cm 0.5cm 1.8cm 2.5cm},clip]{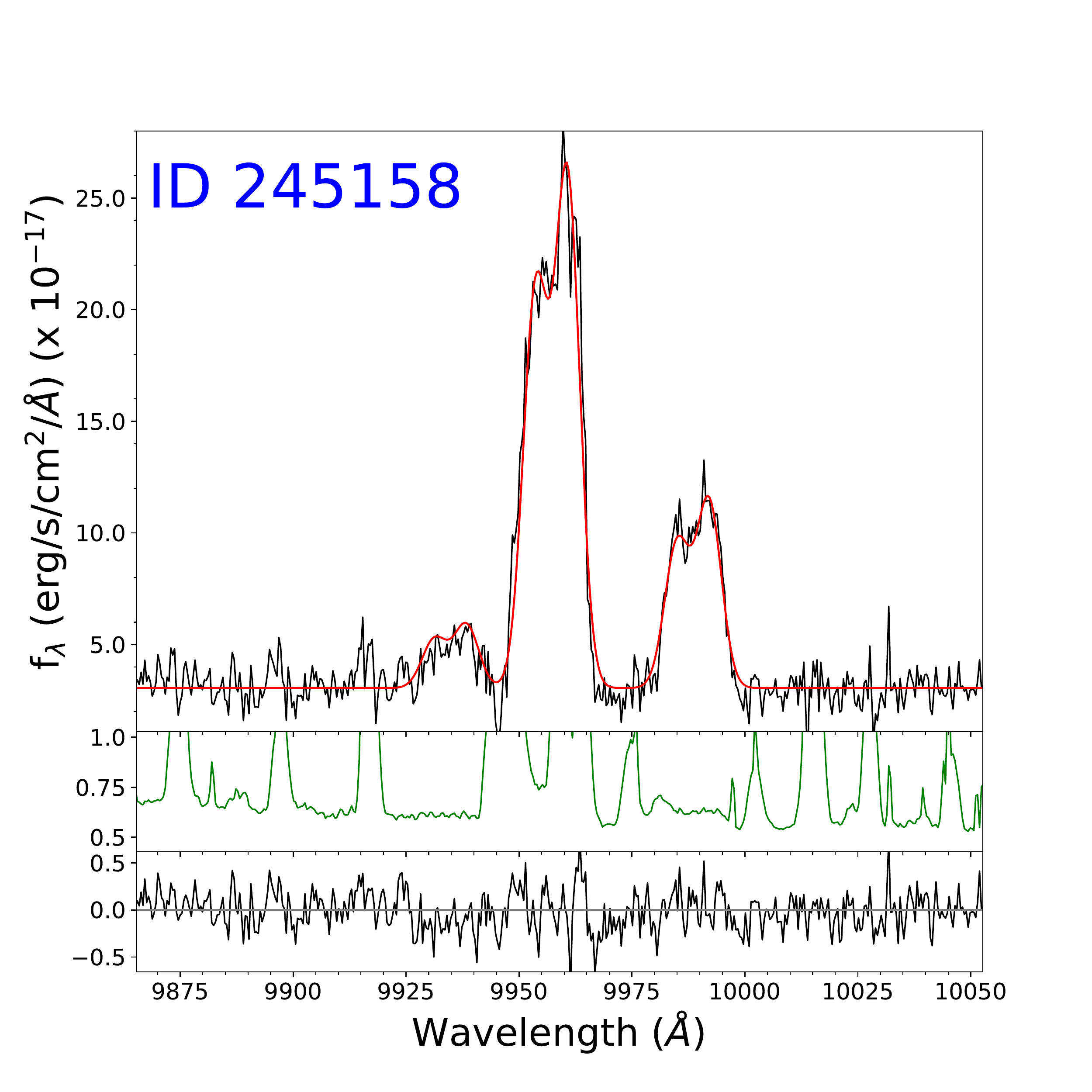}\hspace*{-0.75em}
  \includegraphics[angle=0,height=4.9cm,trim={1.89cm 0.5cm 1.8cm 2.5cm},clip]{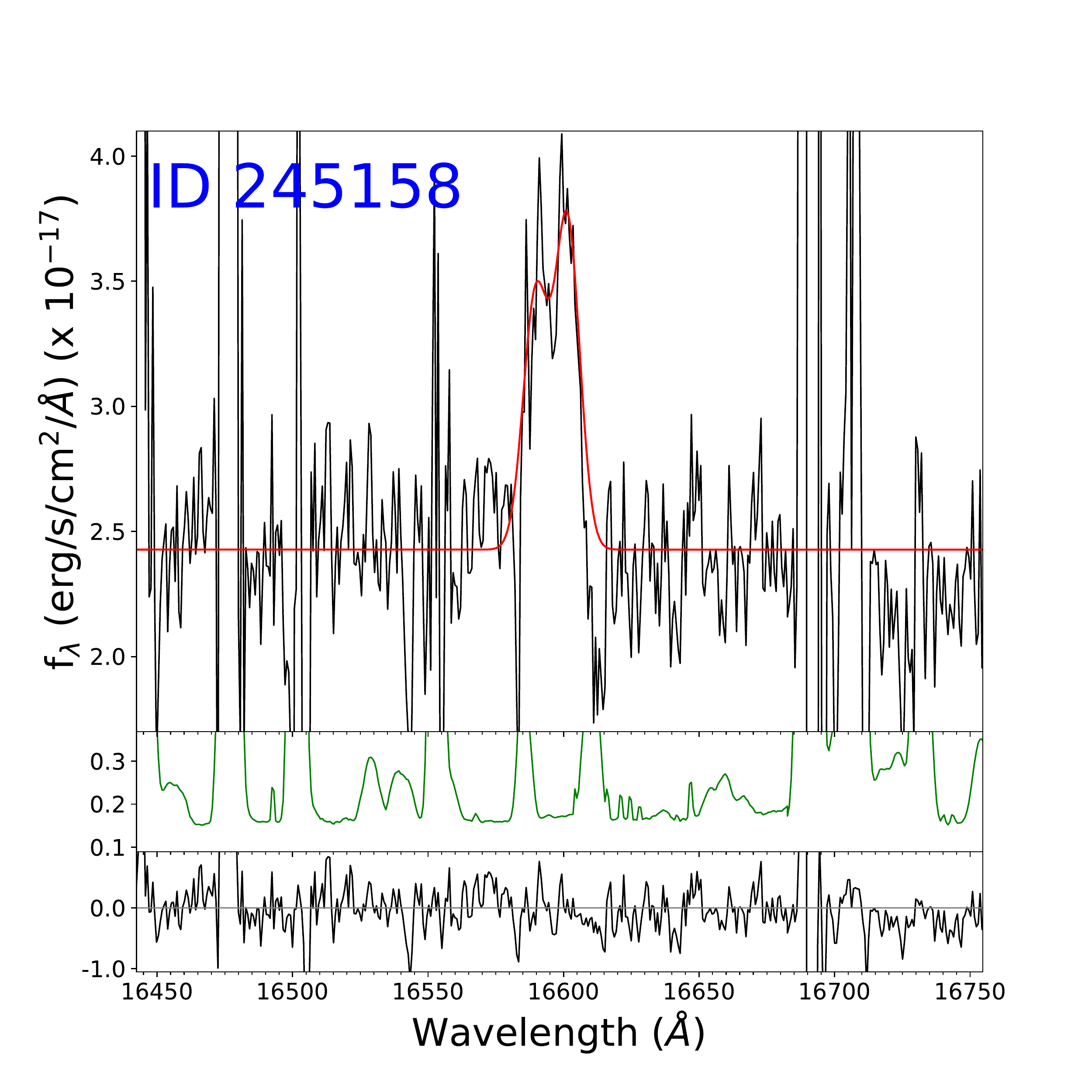}\hspace*{-0.75em}
  \includegraphics[angle=0,height=4.9cm,trim={1.89cm 0.5cm 1.8cm 2.5cm},clip]{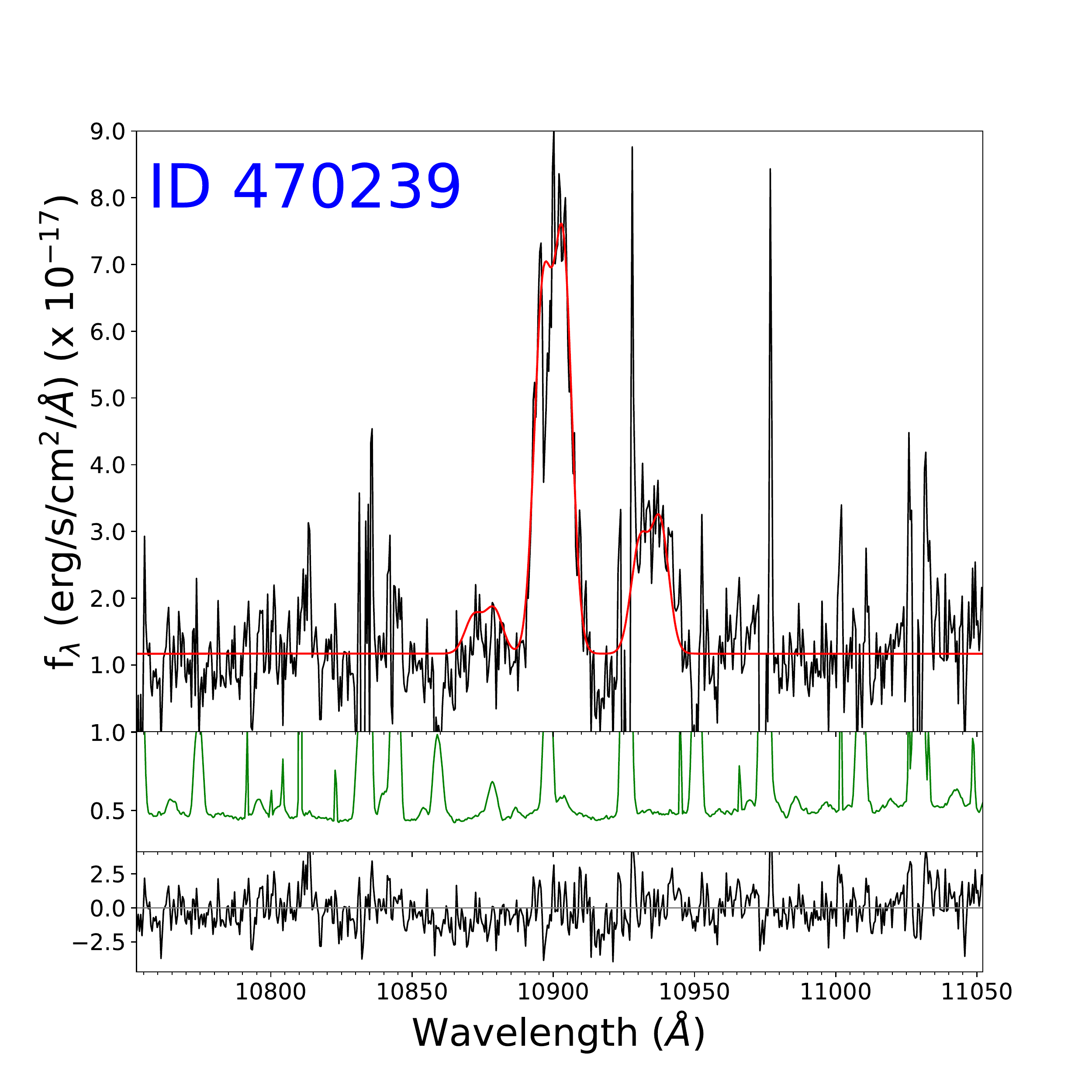}\hspace*{-0.75em}
  \includegraphics[angle=0,height=4.9cm,trim={1.89cm 0.5cm 1.8cm 2.5cm},clip]{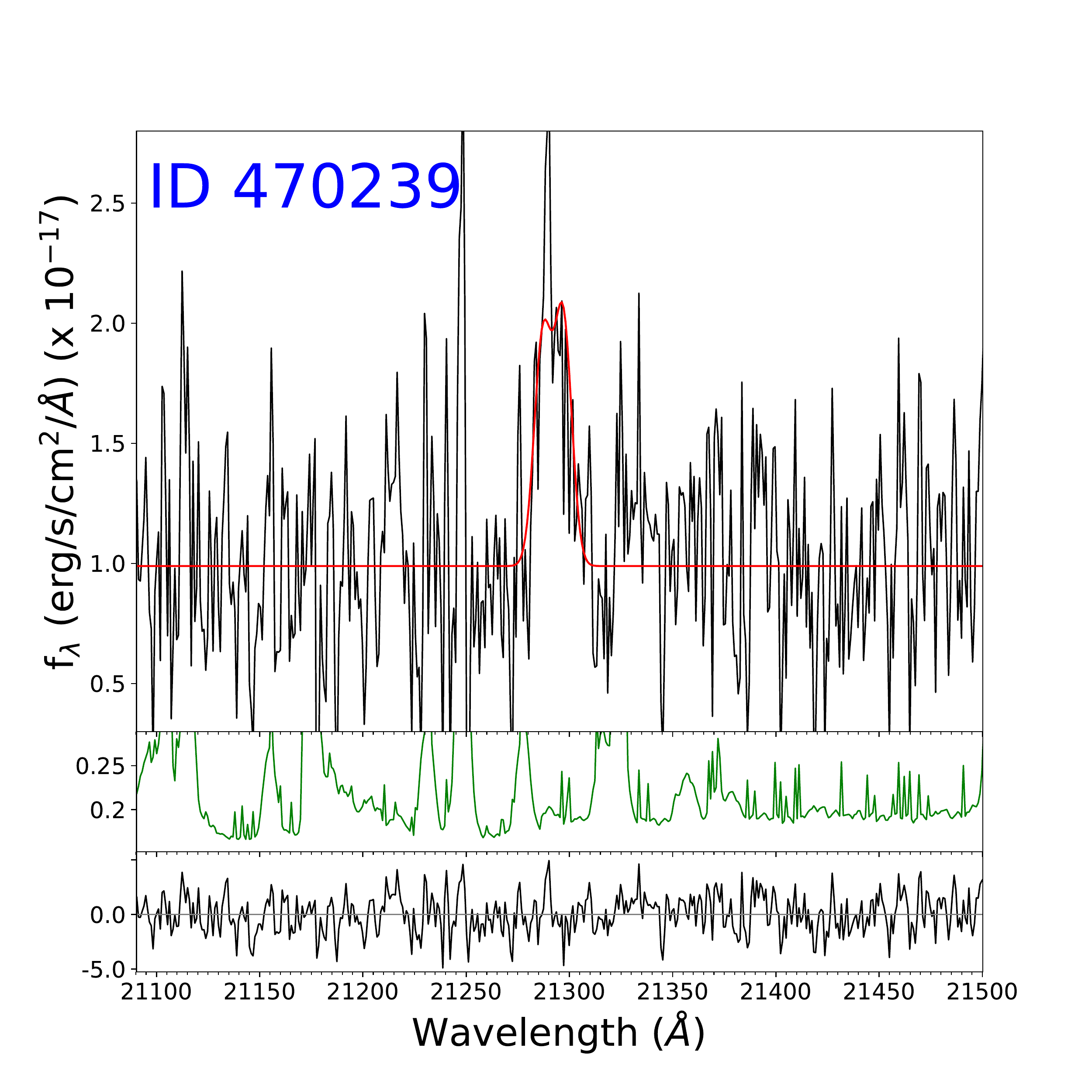}\hspace*{-0.75em}
  
  \hspace*{-2em}
  \includegraphics[angle=0,height=4.9cm,trim={0.5cm 0.5cm 1.8cm 2.5cm},clip]{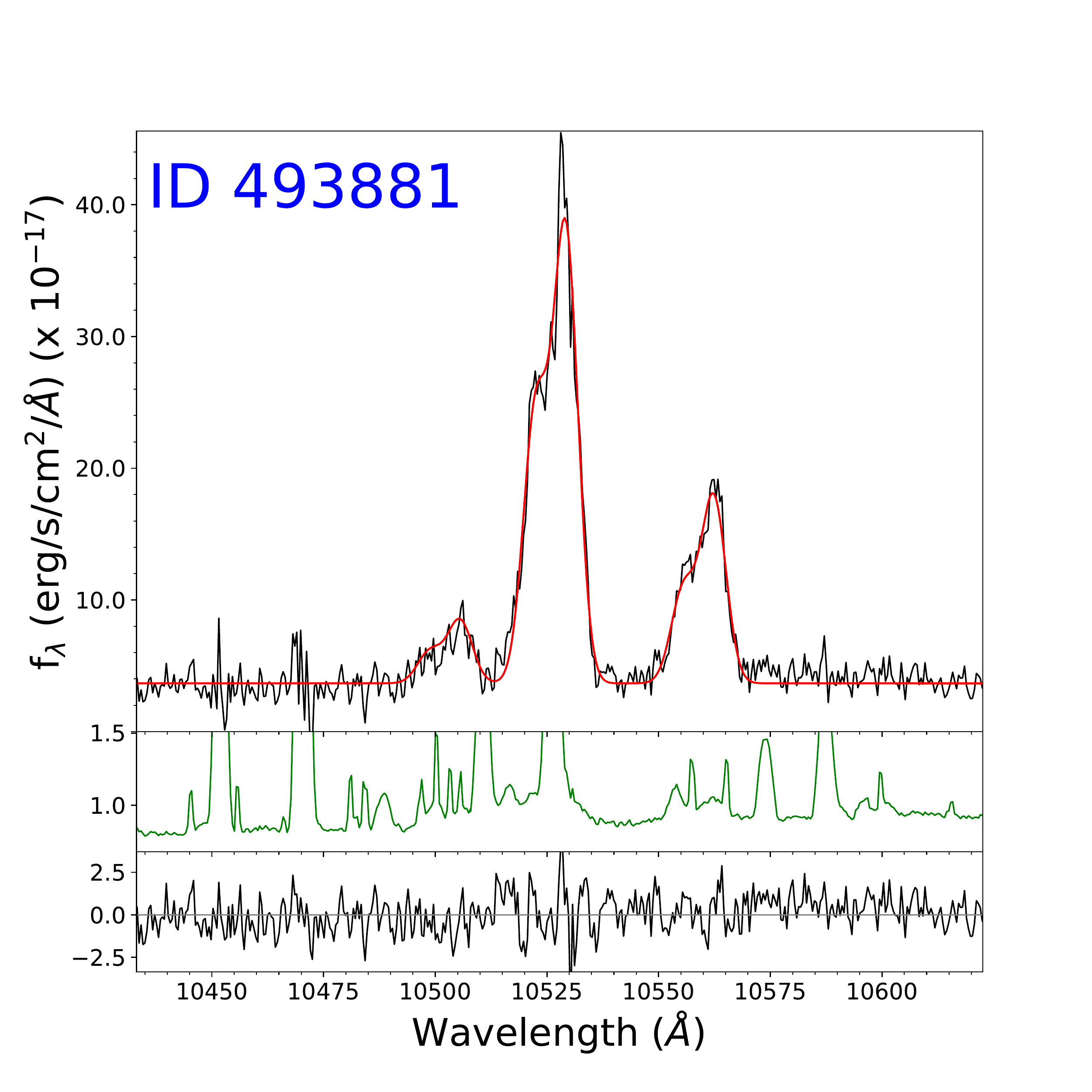}\hspace*{-0.75em}
  \includegraphics[angle=0,height=4.9cm,trim={1.89cm 0.5cm 1.8cm 2.5cm},clip]{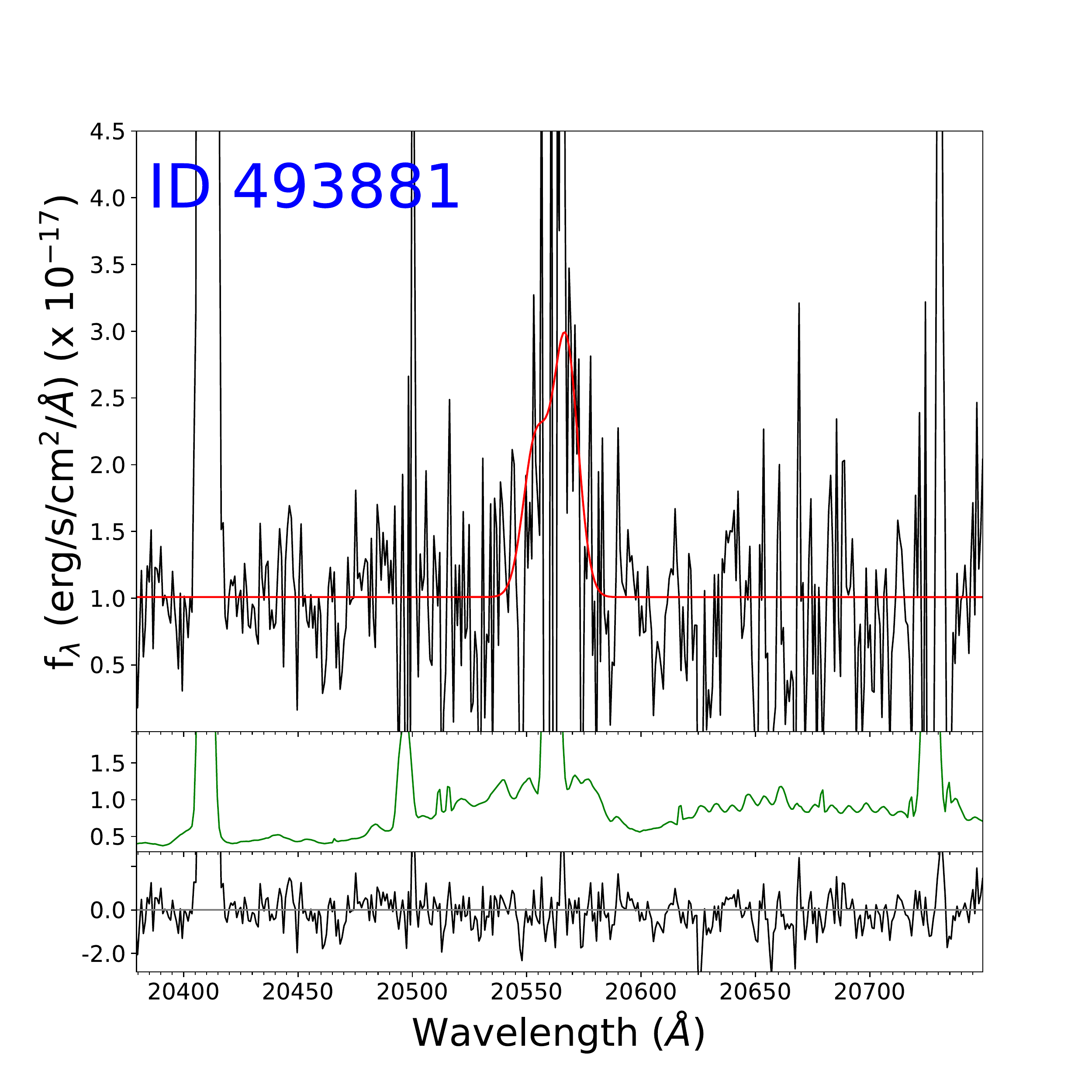}\hspace*{-0.75em}
  \includegraphics[angle=0,height=4.9cm,trim={1.89cm 0.5cm 1.8cm 2.5cm},clip]{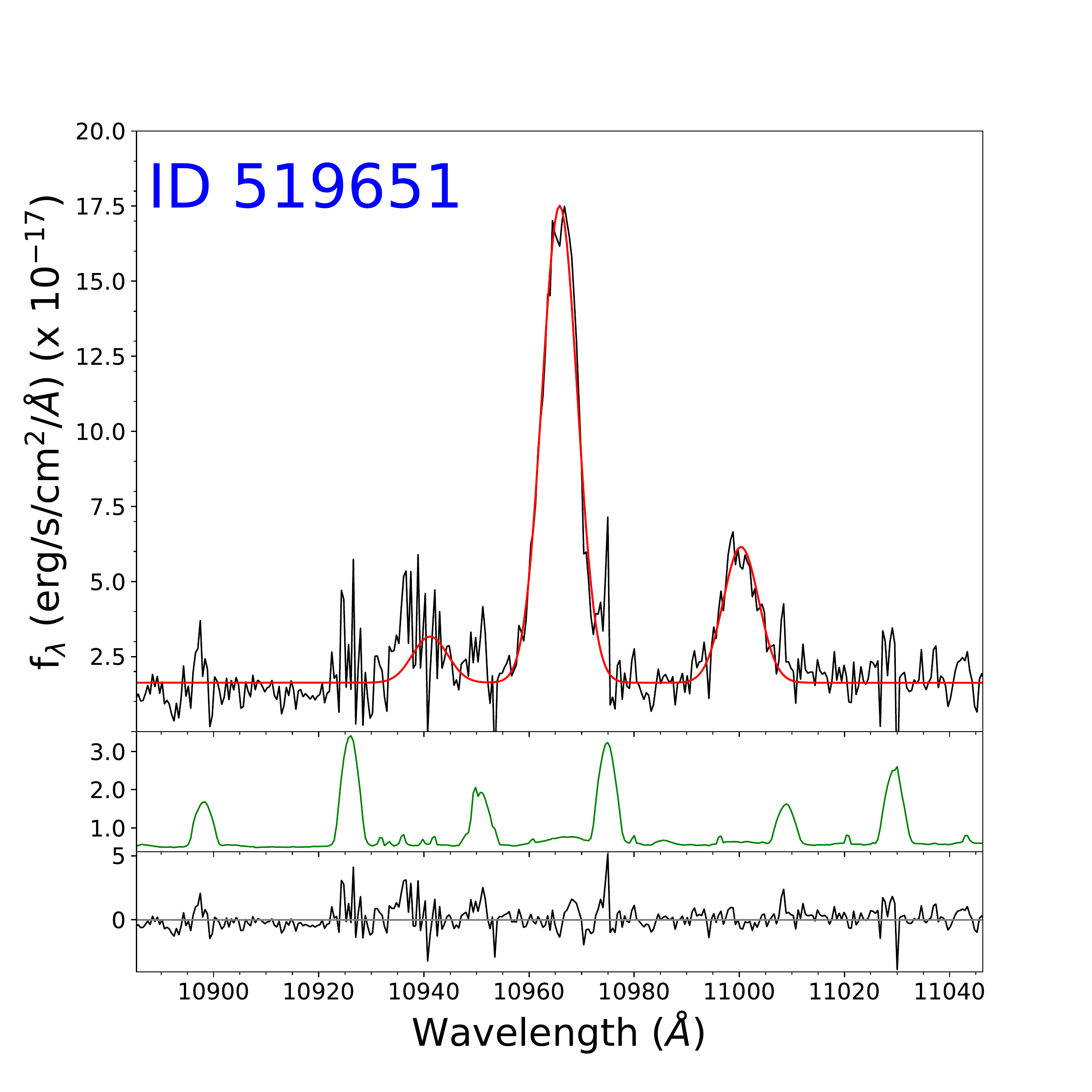}\hspace*{-0.75em}
  \includegraphics[angle=0,height=4.9cm,trim={1.89cm 0.5cm 1.8cm 2.5cm},clip]{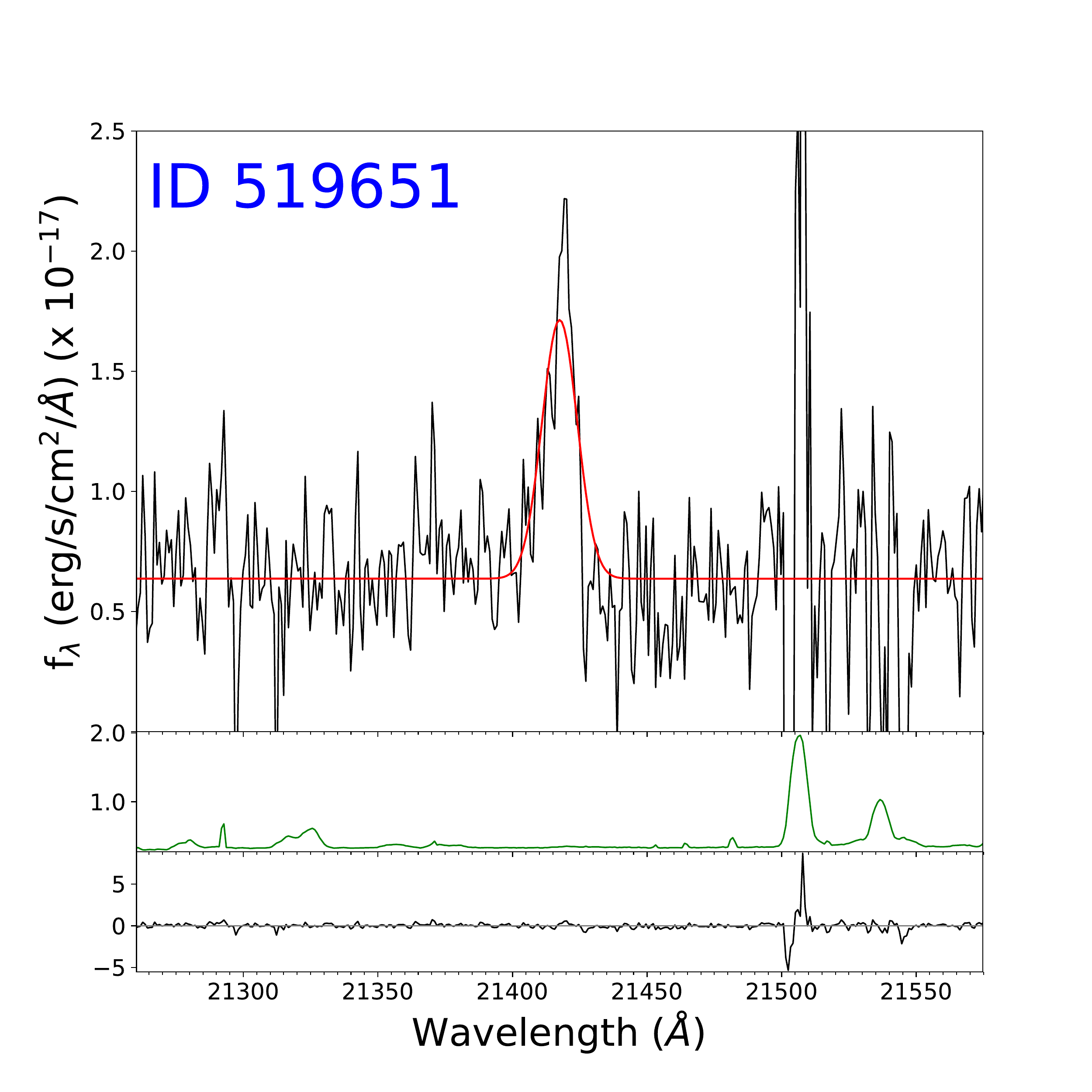}\hspace*{-0.75em}
  
  \caption{\small H$\alpha$ and Pa$\beta$ (Pa$\gamma$ for ID $245158$) emission lines for 4 representative galaxies. In each panel, the spectra around the lines are shown, together with their best-fit gaussians in red, derived with MPFIT \citep{markwardt09}. In the bottom panels we display the noise and the residual of the fit (data-model) normalized by the noise.}\label{figurelines}
\end{figure*}

FIRE is a single slit near-infrared spectrometer mounted at the Magellan $6.5m$ Baade Telescope, covering the wavelength range $0.82$-$2.4\mu m$. We observed in the cross-dispersed echelle mode, choosing a slit width of $1''$ to maximize the incoming light from our targets. This configuration provides a spectral resolution R $\simeq 3000$, which helps  reducing the effect of OH sky-emission. We refer to \citet{simcoe13} for a complete description of the instrument and its performances. 

Our observations were performed in two runs during the nights of $17$-$18$ March 2017 and $22$-$23$ March 2018. We prioritized targets for observations based on two criteria: (1) the presence of a nearby ($\lesssim 30''$ from the target) bright (J $< 19$-$20$ mag) star to facilitate acquisition, and (2) maximization of the ratio \SFRIR/D$_L^2$(z), where D$_L$ is the luminosity distance. The latter condition select galaxies with intrinsically brightest emission lines, and tend to bias our observed sample towards the most massive objects (Fig. \cref{selection}). We observed  $11$ targets during the first run and $14$ in the second, for a total of 25 starbursts. Integration times ranged between $30$ and $80$ minutes, longer for galaxies with lower H$\alpha$ S/N  (from real-time reductions), to improve detection of fainter lines. 

The spectra were reduced using the FIRE pipeline \citep{gagne15}. Full details will be given in a forthcoming paper (Calabr{\`o} et al. in prep), also presenting science results for the complete range of observed emission lines.
In Fig. \cref{figurelines}, we show examples of H$\alpha$ and Pa$\beta$ (or Pa$\gamma$) for some of the galaxies with good detection (S/N $>5$) of both lines. Double Gaussian components were fitted to line profiles whenever single Gaussian fits could be rejected based on $\chi^2$ statistics,
resulting always in good fits ($\chi_{red}^2<1.5$). We attribute these double Gaussians to either rotation or the presence of physically separated components.


The target list with main physical properties are presented in Table \ref{Table1}, while i-band, H-band and radio ($3$ GHz) cutout images of representative targets are shown in Fig. \ref{cutouts}. 

\section{Results}\label{results}

\begin{figure*}[t!]
    \centering
    \includegraphics[angle=0,width=18cm,trim={5.4cm 0cm 4.cm 0cm},clip]{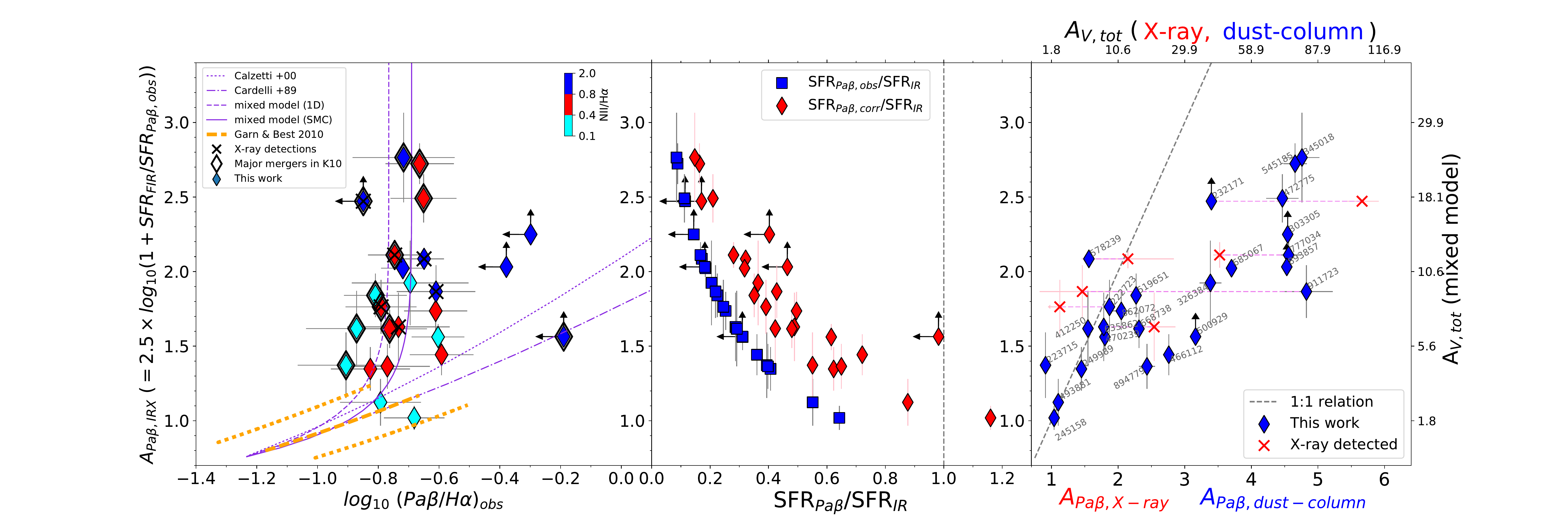}
    \caption{\small \textit{Left:} Diagram comparing the observed Pa$\beta$/H$\alpha$ ratio and A$_{Pa\beta,IRX}$. Upper and lower limits are shown with arrows for four galaxies in the sample, while a color coding highlights their [\ion{N}{2}]/H$\alpha$ values. \textit{Center:} Ratio of the SFR (relative to IR) derived from the observed Pa$\beta$ line (blue) and after correcting the Pa$\beta$ fluxes using the Balmer(H$\alpha$)-Paschen($\beta$) decrement and a \citet{calzetti00} attenuation law (red). We remark that the unobscured UV SFR represents $\sim1\%$ of that derived from the IR, thus its contribution to the total SFR is negligible for our sample. \textit{Right:} Comparison between A$_{Pa\beta,IRX}$, which directly translates into the total A$_V$ towards the center of a mixed model (right axis) with: (\textit{blue circles}) A$_{Pa\beta}$ derived from the dust column-density and (\textit{red crosses}) A$_{Pa\beta}$ calculated from X-ray hydrogen column-density N$_H$ \citep{lanzuisi17} for X-ray detected galaxies, as explained in the text. 
    }\label{calabro}
\end{figure*}

The wide spectral coverage of FIRE and the wealth of photometric data available for our targets makes this a unique sample to investigate attenuation through the use of different indicators as emission lines and the total infrared luminosity. In Fig. \cref{calabro}-\textit{left}\footnote{This plot is equivalent to an IRX-$\beta$ plot \citep{meurer99}.} we compare the ratio of H$\alpha$ and Pa$\beta$ (Paschen-Balmer decrement) to the ratio of SFRs derived from the observed Pa$\beta$ and bolometric IR (A$_{Pa\beta,\text{IRX}}$ $=2.5 \times log_{10}(1+\text{SFR}_{\text{IR}}/\text{SFR}_{Pa\beta , \text{obs} } $)\footnote{For three galaxies in our sample where Pa$\beta$ falls in nearly opaque atmospheric spectral regions or out of FIRE coverage, we use Pa$\gamma$ line to infer the attenuation, estimating Pa$\beta$ flux as $2.2 \times$ Pa$\gamma$ (Table \ref{Table1}). Indeed, both in a mixed model and foreground dust-screen geometry, the expected observed ratio Pa$\beta$/Pa$\gamma$ ranges between $2.1$ and $2.3$, for all the attenuation values in our range.}, where $\text{SFR}_{Pa\beta , \text{obs} }$ has been derived from the observed Pa$\beta$ luminosity, adopting an intrinsic ratio Pa$\beta$/H$\alpha=0.057$ and a standard \citet{kennicutt94} calibration, valid for case B recombination and T$_e=10^4K$.
We show that for our $25$ starbursts, these two ratios, both independent measures of attenuation, do not generally scale as predicted by the \citet{calzetti00} and \citet{cardelli89} attenuation curves\footnote{The Cardelli relation  is actually an extinction law.}. The value of Pa$\beta$/H$\alpha$ rather saturates at $\sim 0.18$  (with a dispersion of $\sim0.08$ dex), qualitatively consistent with an optically thick 'mixed model', in which different lines probe different optical depths. 

As opposed to the foreground dust-screen, a mixed model is made of a uniform extended distribution of young stars and dust inside a volume. In the one-dimensional case, a simple analytic relation can be derived between the observed and intrinsic SFR by integrating along a segment the luminosity contribution from each differential volume element, subject to the extinction of the full optical depth in front of it. This yields:
\begin{equation}\label{eq1}
\frac{SFR_{obs}(\lambda)}{SFR_{intr}(\lambda)} = \frac{L(\lambda)_{obs}}{L(\lambda)_{intr}} = \frac{log_{10}(e)}{0.4} \times \left(\frac{1-10^{-0.8 A_{\text{tot}}(\lambda)}}{2 A_{\text{tot}}(\lambda)}\right)
\end{equation}
where L($\lambda$) is the luminosity of a line at a wavelength $\lambda$ and  $A_{\text{tot}}(\lambda)$ is the total absolute attenuation at $\lambda$ towards the center defined as $k(\lambda)$A$_\text{V,tot}$/$R_V$. 
In the last expression, $k(\lambda)$ and $R_V$ correspond to the local extinction, for which we assumed two extreme cases of a \citet{cardelli89} and an SMC \citep{bouchet85} law, yielding an asymptotic Pa$\beta$/H$\alpha$ ratio of $0.17$ and $0.2$, respectively. 
Using equation \cref{eq1}\footnote{\citet{calzetti94} derived a similar equation (n.19) for a mixing geometry}, we can predict the observed fluxes at all wavelengths as a function of a single parameter,  A$_\text{V,tot}$. For small values of A$_\text{V,tot}$ this model coincides with the standard attenuation curves adopted. For large A$_\text{V,tot}$, the local extinction inside the starburst core increases toward the center until the photons are not able to escape anymore from the galaxy, and are fully absorbed by the outer layers of dust. This leads us to depict heavily obscured starbursts as made of a central optically thick core, invisible to us, and a surrounding {\em skin}, producing the observed optical and near-IR nebular lines.

This picture naturally explains both the larger attenuation and SFR fraction that can be recovered by near-IR observations with respect to optical studies \citep{puglisi17}, as near-IR wavelengths allow us to penetrate deeper in the system.  
Because only the less attenuated light from the skin comes out, from Pa$\beta$ we can recover, on average, 30\% of the total IR SFR (Fig. \cref{calabro}-\textit{middle}). 
However, inside the skin the optical depth becomes quickly large, with median A$_{V,\text{tot, mixed model}}=9$, corresponding to a suppression of $\times4000$ of V-band light from the starburst core centers and up to extreme cases with  A$_{V,\text{tot,mixed model}}\sim30$ (10$^{12}$ in linear scale). Hence, we cannot directly see the starburst cores in the optical/near-IR. 

\section{Discussion}

Can we conclude that z $\sim0.7$ SBs contain extremely obscured cores that are well described by mixed stars/dust models? It is worth considering alternative explanations.
It might be possible that the UV radiation from newly born massive stars is absorbed by dust within HII regions, 
before reaching to ionize HI outside. While strong stellar winds push the dust away to form a screen \citep{calzetti00}, a substantial amount of absorbing dust may still be trapped in the ongoing SF site \citep[][]{caplan86,bell01} particularly in these very dust rich galaxies. This would {\em simulate} the existence of an optically thick SB core, just reducing the fraction of photons seen by HI. This could still represent a mixed model scenario, with mixing occurring at smaller scales. Whether this is a viable option depends on geometry and is difficult to model in detail. 

Alternatively, the discrepant SFRs (coming from the lines and IR bolometric luminosity) may be due to time-variation effects, as L$_{\text{IR}}$ probes longer SFR timescales than emission lines, due to the energy contribution of longer-lived B-type stars  to ionizing O stars. In our case, this would require that most SBs have recently experienced a severe SFR truncation, which seems unlikely. Instantaneous and dust-free SFR tracers, like e.g., through CO[5-4] lines \citep{daddi15}, would shed light and help addressing definitely  this possibility. 

Due to the optically thick  cores, the mixed model also implies that it might not be possible to detect AGNs from optical and near-IR observations, if located in the coalescing center. 
Interestingly, 
Fig. \cref{calabro} shows that our SB galaxies often display high  N2 ($\equiv \log$[\ion{N}{2}]/H$\alpha$) indexes, which might suggest highly ionizing, AGN-dominated emission. We also see a correlation, significant at $>95$\% confidence level (Spearman correlation coefficient $r=0.5$), between N2 and A$_{V,\text{tot, mixed model}}$\footnote{This incidentally suggests that the four galaxies with Pa$\beta$ upper limits are also very highly obscured,  having relatively high N2.}. 
We argue that instead, in the majority of our targets with enhanced [\ion{N}{2}]/H$\alpha$ (and relatively higher obscurations), the line emission may be driven by shocks, which were already shown to contribute up to $50\%$ in local ULIRGs in  latest merger stages   \citep{rich15}. In case of shock contribution, the attenuations that we have inferred through A$_{Pa\beta,IRX}$ would represent  lower limits, but the line ratios will not be affected as Case~B recombination regime 
still holds.

Nevertheless, it would be crucial to obtain independent estimates of actual attenuations towards the  cores. 
One possibility is provided by AGNs.  
We searched for evidence of AGNs among our SBs using multiple dust-free multi-wavelength tracers in the radio and X-rays. While none of our SBs show significant radio excess, either following the criteria of \citet{delmoro13} and the less stringent requirements of \citet{liu18} (all assuming an IR-radio correlation), six galaxies (ID $578239$, $635862$, $777034$, $232171$, $222723$, and $911723$) are detected by XMM-Newton, Chandra or NuStar \citep[][]{cappelluti09,marchesi16,civano15} with luminosities much higher than what expected from their SFRs \citep[][]{ranalli04}. 
The same objects are also the only ones in which we detect a mid-IR dusty torus component through SED fitting (only tentative for SB $578239$). 
\textcolor{black}{Their X-ray hardness ratios  were converted in obscuring column densities (N$_H$) by Lanzuisi et al. (2017; see Fig.19 in La~Massa et al 2016 for the method), which are N$_H$ upper limits for  gas/dust obscurations to the cores (part of the obscuration would happen within the torus itself).}
The relation of \citet{guver09} allows to convert N$_H$ into a total A$_V$ and  (N$_H$(cm$^{−2}$) = $2.21 \times 10^{21}$ A$_V$(mag)). This returns generally very high X-ray obscurations for the AGNs (Fig. \cref{calabro}-\textit{right}), supporting the presence of high obscuration in their center, as required by the mixed model.

As a further check, we computed the column density of gas in the starbursts cores using the total molecular mass M$_{\text{mol}}$, inferred as M$_{\text{mol}}= 8.05 + 0.81 \times$ log(SFR) \citep[][, assuming conservatively the starburst case]{sargent14} and the radio size, measured with GALFIT by fitting a gaussian profile (convolved with the PSF) to their VLA (3 GHz) images (Fig. \cref{cutouts}). 
From Fig. \cref{calabro}-\textit{right} we can see that for half of our sample, at relatively low-moderate obscurations within the probed range, the attenuation inferred from this method is consistent with the 
mixed model. On the other hand, towards the highest obscurations, this approach suggests  even larger attenuations. 
Some fraction of the  emission line fluxes might come from foreground regions unrelated to the starbursting cores, presumably residual material from the merging galaxies. Accounting for this extra, modestly attenuated component would result in substantial increase of the A$_{\text{V,tot}}$ of the starburst core, providing better agreement with these column densities estimates. A proper correction for this effect would require IFU observations, that could  also clarify whether our objects are similar to local ULIRGs, for which higher values of nuclear attenuations (A$_{\text{V}}\sim10$-$1000$ mag) are reported in several studies \citep[e.g.,][]{scoville98,genzel98}.
All in all, it appears that very heavily embedded cores are indeed present in these galaxies. 

What is the origin of these extremely obscured cores? Unsurprisingly (Fig.~2), morphological classification was independently derived for $18$ starbursts in our sample \citep[][see Tab.~1]{kartaltepe10}, and over 83\% of them were identified as mergers: $11$ as major mergers ($61\% $), showing distorted or double nuclei, tidal tales, bridges or overlapping disks, and $4$ (22\%) as minor mergers, characterized by at least slightly disturbed morphology (e.g., warped disks, asymmetric spiral arms, small companion at same z, etc).  
Visual inspection for the remaining sources  suggest that merger origin is at least plausible for the vast majority of our sample.
Mergers are, in fact, more commonly identified among less obscured systems (Tab.~1), which is understandable given that in the later coalescence phases any remaining merger signature becomes subtle
(see, e.g., two of the three morphologically non-merger objects in our sample classified as Ellipticals/S0 by \citet{kartaltepe10}, and the three of them have A$_{\text{V,tot}}>9$).
It is thus tempting to attribute the large range of observed properties apparently defining  a sequence of obscurations as reflecting different merger phases, to varying progenitor properties (including, e.g., the gas fraction of merging galaxies and the impact geometry), or a combination of them. Nevertheless, we cannot definitely exclude with our data that the sequence may be also reflecting the amount of foreground contamination, thus unrelated to the real obscuration of the core.

Our work suggests that deeply embedded merger events still largely dominate among sample of SBs galaxies at least to $z\sim0.7$, which corresponds to 6.3~Gyr lookback time, an epoch with  galaxy specific SFRs $>5\times$ larger than local on average. 
At even higher redshifts, it becomes much harder to identify mergers
from their morphological signatures due to surface brightness dimming and widespread presence of clumpy/irregular galaxies.  We argue instead that higher-z mergers might be even more efficiently identified searching for evidences of extreme levels of obscurations, given our results and also consistently with simulations \citep[e.g.][]{dimatteo05},  representing a clear footprint of their origin. In fact, we are not aware of any viable alternative mechanism that could produce galaxy-wide obscurations of 10+~mag in the V-band: normal disk-like galaxies display much lower obscurations (orange lines in Fig.4-left). Near-IR rest-frame spectra of galaxies will be easily accessible soon with JWST up to z $\simeq 7$ and down to much fainter levels, and will allow testing and applying this idea.

\begin{table*}[t!]
\centering
\begin{threeparttable}
\caption{{Main properties of the Magellan starbursts}}
\label{Table1}
\smallskip
\setlength\tabcolsep{8.pt}
\begin{tabular}{p{4.5mm}p{9mm}p{6mm}p{4mm}p{6mm}p{16.5mm}p{16mm}p{16mm}p{13mm}p{16mm}p{4mm}}
\hline\hline
\textbf{ID}	& \textbf{RA} & \textbf{DEC} & \textbf{z$_{\textbf{spec}}$}	& \textbf{log(M$_\ast$)} & \textbf{log(LIR$_\text{\textbf{SFR}}$)} & \textbf{H$\boldsymbol{\alpha}$} & \textbf{Pa$\boldsymbol{\beta}$} & \textbf{A$_{\textbf{V,tot}}$} & \textbf{N$_{\textbf{H,X}}$} & \textbf{M$_{type}$} \\
 & (deg) & (deg) & & (M$_\odot$)  & (L$_\odot$) & (\scriptsize{$10^{-17} \frac{erg}{s cm^2}$}) & (\scriptsize{$10^{-17} \frac{erg}{s cm^2}$}) & (mag) & (cm$^{-2}$) & \\
\hline\hline
245158 & 150.18854 & 1.65498 & 0.5172 & 10.7 & 11.89 $\pm$ 0.07 & 249.6 $\pm$ 30.1 & 52.0 $\pm$ 5.6\textsuperscript{\textdaggerdbl} & 1.9 $\pm$ 0.1 & - & S,m \\
493881 & 150.74967 & 2.04707 & 0.6039 & 10.8 & 12.09 $\pm$ 0.06 & 330.9 $\pm$ 30.2 & 53.3 $\pm$ 11.5                                & 2.7 $\pm$ 0.2 & - & - \\
223715 & 149.76537 & 1.61702 & 0.5174 & 10.7 & 11.78 $\pm$ 0.03 & 204.0 $\pm$ 20.8 & 28.5 $\pm$ 6.6                                 & 3.6 $\pm$ 0.2 & - & m,S \\
249989 & 150.68540 & 1.66108 & 0.6656 & 10.6 & 11.90 $\pm$ 0.08 & 127.6 $\pm$ 15.9 & 19.0 $\pm$ 3.3                                 & 4.4 $\pm$ 0.2 & - & MIII \\
894779 & 150.42710 & 2.65644 & 0.5506 & 10.0 & 11.79 $\pm$ 0.03 & 129.7 $\pm$ 19.3 & 22.0 $\pm$ 3.8                                 & 4.5 $\pm$ 0.2 & - & MIV \\
466112 & 149.99928 & 2.00599 & 0.7607 & 10.3 & 12.16 $\pm$ 0.04 & 81.7  $\pm$ 6.2  & 20.9 $\pm$ 3.5                                 & 5.2 $\pm$ 0.3 & - & MIII \\
470239 & 150.48155 & 2.01096 & 0.6609 & 10.6 & 12.12 $\pm$ 0.17 & 92.3  $\pm$ 9.5  & 23.0	$\pm$ 2.2                               & 6.2 $\pm$ 0.2 & - & - \\
500929 & 149.76844 & 2.05935 & 0.9498 & 10.8 & 12.25 $\pm$ 0.14 & 21.9  $\pm$ 5.3  &  $<$ 14.3\textsuperscript{\textdaggerdbl}      & $>$ 6.2 & - & MIII \\
412250 & 150.74171 & 1.91764 & 0.8397 & 10.3 & 12.21 $\pm$ 0.04 & 118.6 $\pm$ 10.5 & 16.0 $\pm$ 4.7                                 & 6.7 $\pm$ 0.6 & - & MIII \\
668738 & 150.21020 & 2.31168 & 0.7481 & 10.8 & 12.20 $\pm$ 0.04 & 81.4  $\pm$ 7.5  & 14.1 $\pm$ 1.4                                 & 6.7 $\pm$ 0.2 & - & MIII \\
635862 & 149.69589 & 2.26450 & 0.5508 & 11.0 & 11.58 $\pm$ 0.10 & 59.7  $\pm$ 8.8  & 11.0	$\pm$ 2.7                               & 6.8 $\pm$ 0.6 & 22.54 $\pm$ 0.14 & - \\
862072 & 150.12329 & 2.60376 & 0.6811 & 11.1 & 11.98 $\pm$ 0.07 & 58.3  $\pm$ 6.0  & 14.3 $\pm$ 1.9                                 & 7.8 $\pm$ 0.4 & - & m,S \\
222723 & 150.17321 & 1.61632 & 0.5254 & 11.0 & 12.05 $\pm$ 0.05 & 150.7 $\pm$ 11.1 & 24.4 $\pm$ 4.8                                 & 8.1 $\pm$ 0.5 & $<$ 21.66 & MV \\
519651 & 150.43020 & 2.08688 & 0.6709 & 10.5 & 12.15 $\pm$ 0.05 & 126.0 $\pm$ 10.5 & 19.5 $\pm$ 3.0                                 & 8.8 $\pm$ 0.5 & - & MIV \\
911723 & 149.68134 & 2.68108 & 0.6606 & 10.8 & 12.03 $\pm$ 0.02 & 65.9  $\pm$ 7.2  & 16.2 $\pm$ 3.1                                 & 9.1 $\pm$ 0.6 & 21.98 $\pm$ 0.36 & E \\
326384 & 149.51786 & 1.78357 & 0.8042 & 10.3 & 12.20 $\pm$ 0.07 & 68.8  $\pm$ 8.8  & 12.4 $\pm$ 3.8                                 & 9.8 $\pm$ 1.1 & - & S \\
685067 & 149.74730 & 2.34574 & 0.3735 & 11.0 & 11.75 $\pm$ 0.01 & 136.9 $\pm$ 16.5 & 26.1 $\pm$ 1.2                                 & 10.9 $\pm$ 0.3 & - & - \\
893857 & 150.15995 & 2.65434 & 0.8512 & 11.1 & 12.26 $\pm$ 0.09 & 27.4  $\pm$ 1.6  & $<$ 11.5\textsuperscript{\textdaggerdbl}       & $>$ 11 & - & E \\
578239 & 150.76543 & 2.18099 & 0.5578 & 11.1 & 12.21 $\pm$ 0.23 & 116.5 $\pm$ 10.4 & 26.1 $\pm$ 1.6                                 & 11.7 $\pm$ 0.3 & 22.36 $\pm$ 0.31 & - \\
777034 & 150.15025 & 2.47517 & 0.6889 & 10.8 & 12.43 $\pm$ 0.08 & 114.0 $\pm$ 13.3 & 20.5 $\pm$ 1.8                                 & 12.0 $\pm$ 0.4 & 22.96 $\pm$ 0.04 & MIV \\
303305 & 150.48305 & 1.74796 & 0.5306 & 10.7 & 11.84 $\pm$ 0.10 & 19.9  $\pm$ 3.6  &  $<$ 10.0\textsuperscript{\textdaggerdbl}      & $>$ 13.9 & - & - \\
232171 & 150.06033 & 1.63269 & 0.5251 & 11.1 & 11.71 $\pm$ 0.02 & 60.1  $\pm$ 14.2 & $<$ 8.5                                        & $>$ 17.6 & 23.83 $\pm$ 0.1 & MII \\
472775 & 150.48148 & 2.01362 & 0.6604 & 10.8 & 12.17 $\pm$ 0.06 & 44.3  $\pm$ 4.1  &  9.9 $\pm$ 1.6                                 & 18.0 $\pm$ 1.3 & - & - \\
545185 & 149.52802 & 2.12725 & 0.5337 & 10.4 & 12.10 $\pm$ 0.06 & 49.8  $\pm$ 4.2  & 10.9	$\pm$ 0.9                               & 22.5 $\pm$ 1.0 & - & MIII \\
345018 & 149.72556 & 1.81069 & 0.7521 & 10.6 & 12.25 $\pm$ 0.06 & 37.0  $\pm$ 3.6  &  7.1 $\pm$ 2.1                                 & 23.7 $\pm$ 3.4 & - & m,S \\
\hline
\end{tabular}

\begin{tablenotes}\footnotesize
\item[] \textbf{Note.} ID, RA, DEC (J2000) and M$_\ast$ are from \citet{laigle16}. The $1$-$\sigma$ error on M$_\ast$ is $0.1$ dex. Line fluxes are measured from aperture corrected spectra. Aperture correction errors ($\sim 0.04$ dex) are included in the uncertainties. The infrared luminosities (integrated between $8$-$1000$ $\mu m$), are AGN-torus decontaminated. H$\alpha$ and Pa$\beta$ are corrected for stellar absorption, assuming EW$_{abs}=$ $2.6$ and $2$ \AA, respectively.
The morphological type (M$_{type}$) of \citet{kartaltepe10}: E=Elliptical/S0; S=spiral/disc; m=Minor merger; M=Major merger (I:first approach, II:first contact, III:pre-merger, IV:Merger, V:Old merger/merger remnant). The galaxies are ordered with increasing A$_{\text{V,tot}}$. \textdaggerdbl: derived as Pa$\beta$ $=2.2\ \times$ Pa$\gamma$
\end{tablenotes}
\end{threeparttable}
\smallskip
\end{table*}

\acknowledgments
We thank the referee for useful suggestions, G.Rudie for assistence with Magellan observations, Nicol{\'a}s Ignacio Godoy for data reduction, and Daniela Calzetti for discussions. We acknowledge support from FONDECYT regular programs 1150216 and 1170618, by the Brain
Pool Program,  funded by the Ministry of Science
and ICT through the Korean National Research Foundation, and RadioNet conference funding   
(2018H1D3A2000902).


\end{document}